\documentclass[12pt]{article}
\usepackage[pctex32]{graphics}
\begin{document}
~
\vspace{1cm}
\begin{center} {\Large \bf   Tube of  (Circle F, D0, D2) Bound State in Melvin  Background}
                                                  
\vspace{2cm}

                      Wung-Hong Huang\\
                       Department of Physics\\
                       National Cheng Kung University\\
                       Tainan,Taiwan\\

\end{center}
\vspace{1cm}
\begin{center} {\large \bf  Abstract} \end{center}
By using the Born-Infeld action we show that the $m$ circular fundamental strings, $n$ D2-branes and $k$ D0-branes could become a tubular bound state which is prevented from collapsing by the magnetic force in the Melvin background.   However, if the ratio $m/n$  is larger then a critical value the tube will become unstable and collapse to zero radius.  We make analyses to find the critical value and tube radius therein.  The tube configurations we found are different from the well known tubular bound states of straight fundamental strings, D0 and D2-branes, which are supported by the angular momentum.

\vspace{3cm}
\begin{flushleft}
E-mail:  whhwung@mail.ncku.edu.tw\\
2004/08/14
\end{flushleft}
\newpage
\section  {Introduction}
In an interesting paper [1-3], Mateos and Townsend showed that a bunch of straight fundamental strings with D0-branes can be blown-up to a tubular D2-brane.    The tube configuration is supported against collapsing by the angular momentum generated by the crossed electric and magnetic  Born-Infeld (BI) fields of  the straight fundamental strings and  D0-branes, respectively.   The background in this case is trivial as there is no external force as that in the Myers effect [4] in which the external RR four-form field could expand D0 branes into the fuzzy sphere to stabilize the system. 

In this paper we will use the Born-Infeld action to find a new tubular configuration in the Kaluza-Klein Melvin magnetic tube spacetime [5-8].    We will show that the magnetic force from the Melvin background could prevent 
the tubular bound configurations of  $m$ circular fundamental strings, $n$ D2-branes  and $k$ D0-branes from collapsing.   However, if the ratio $m/n$ is larger then a critical value then the tube will become unstable and collapse to zero radius.  We present a simple analysis to find the critical value and tube radius.  We discuss the physical reasons of how would the tube radius and critical value depend on the numbers of $m$ and $n$.  We also show that tubular bound state of $m$ circular fundamental strings, $n$ D2-branes  and $k$ D0-branes do not exist in the homogeneous magnetic background [7]. 

\section{A Tubular Solution }
The Melvin metric [5] is a solution of  Einstein-Maxwell theory, which describes a static spacetime with a cylindrically symmetric magnetic flux tube.  The ten dimensional  Melvin metric we used is described by [8]
$$ ds_{10}^{2} = \Lambda^{1/2}\left(-dT^2+dy_mdy^m+dZ^2+dR^2\right)
  +\Lambda^{-1/2}R^2d\,{\Phi}^2 , $$
$$  e^{4\phi/3}=\Lambda \equiv 1+f^2R^2 ,~~~~~ A_\Phi=\frac{{\tilde B}R^2}{2\Lambda}, ~~~~~~ \eqno{(2.1)} $$
where the parameter $f$ is the magnetic field along the $z$-axis defined by
$f^2=\frac{1}{2}F_{\mu\nu}F^{\mu\nu}|_{\rho=0}$ and $\phi$ is the dilaton field.   This spacetime, as described in [8],  could be obtained by the reduction from 11d with $A_\Phi$ being RR type IIA vector. 

  The Born-Infeld Lagrangian of tubular bound state of $m$ circular  fundamental strings, $n$ D2-branes  and $k$ D0-branes, for unit tension, is written as [1,10]
$${\cal L} =  - n\,e^{-\phi}~\sqrt{- \det (g +F)}\,,   \eqno{(2.2)} $$
where $g$ is the induced worldvolume 3-metric and $F$ is the BI 2-form
field strength.     If we take the worldvolume coordinates to be $(t,z,{\varphi})$ with ${\varphi} \sim {\varphi} + 2\pi$, then we may fix the worldvolume diffeomorphisms for a tubular topology by  the `physical' gauge choice
$$T=t , ~~~~Z=z,~~~~~ \Phi= \varphi .  \eqno{(2.3)} $$
For a static straight tube solution of circular cross section with radius $R$ the induced metric is
$$ ds^2(g) =  \Lambda^{1/2}\,\left(- dt^2 + dz^2\right)+\Lambda^{-1/2}\,R^2\,d{\varphi}^2 .\eqno{(2.4)} $$
We will allow for a time-independent electric field $E$ and magnetic field $B$ such that the BI 2-form field strength is [1,10]
$${\bf F}= E \, dt\wedge d\varphi + B \, dz\wedge d\varphi . \eqno{(2.5)}$$
Note that the tubular solution we searching are the $n$ static cylindrical D2-branes of radius $R$ binding with $m$ circular F-strings and $k$ D0-branes. The BI 2-form field strength we considered is described in (2.5).   Thus the circular F-strings are fusing inside the D2 worldsheet by converting itself into homogenous electric flux.  As the direct along the electric is a circle with radius $R$\, the open strings now stretch around the circle and the two ends  join to each other with a finite probability [9-11]. 

 It shall be mentioned that the supertubes considered in [1] have the BI 2-form field strength $ {\bf F}= E \, dt\wedge dz + B \, dz\wedge d\varphi $.   Thus the F-strings are along the $z$ axial, which is different from our system.  Note that the EM-fields in [1] will generate an angular momentum to stabilize the tubular D2 brane and prevent its collapsing to zero radius.  Our tubular solutions are  prevented from collapsing by the magnetic force in the Melvin background and therefore are different from those in [1].

Under these conditions the Lagrangian becomes
$$ {\cal L} = -\,{n\,\sqrt {R^2 - E^2 + B^2}~\over ~~\sqrt {1 + f^2R^2}~},  \eqno{(2.6)}$$
and the momentum conjugate to $E$ takes the form

$$\Pi \equiv {\partial{\cal L}\over \partial E}  =  \,{n\,E \over~\sqrt{1 + f^2R^2} ~\sqrt {R^2 - E^2 + B^2}~}.   \eqno{(2.7)}$$ 
\\
The conserved F-string and D0 charges are defined by [1-3,9-11]

$$ m = {1\over 2\pi}\int d \varphi \, \Pi\, ,\eqno{(2.8a)}$$ 
$$ k = {1\over 2\pi}\int d \varphi \,  B\,.\eqno{(2.8b)}$$ 
\\
The corresponding Hamiltonian density per unit length is 
$$ {\cal H}_{(B,\Pi,n,f)}\left(R\right) \equiv \Pi E - {\cal L}  = \sqrt {R^2 + B^2}\, \,\sqrt{{n^2\over 1 + f^2R^2}\, + \Pi^2}.   \eqno{(2.9)}$$
In the case of $f\ne 0$, then for the fixed numbers of $n$, $\Pi$ and $B$, which corresponding to the fixed numbers of  D2-branes, F-strings and D0-branes respectively,  the bound state of tube  has a minimum energy at a finite radius, as shown in figure 1. 
\\
\\
\scalebox{1}{\hspace{4cm}\includegraphics{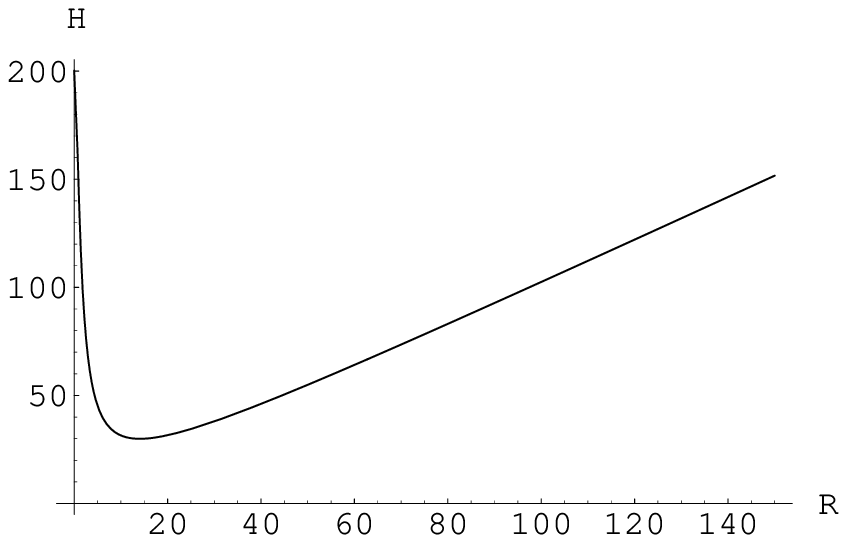}}\\ 
\\
\hspace{2cm}{\it ~~~Figure 1. The function behavior of  ${\cal H}_{(B,\Pi,n,f)}\left(R\right)$ in (2.9) with $(B,\Pi,n,f) = (10,1,20,1)$.  The energy density has a local minimum at finite radius R which is the radius of the tube}\\

To make more analyses about the solution we can from  (2.9) see that at large value of $R$ the energy $H \sim |\Pi\, R|$ \, which is an increasing function of $R$.   Thus, if at small value of $R$ the energy was a decreasing function then there will have a local minimum at finite $R$ which specifies the radius of the tube.   Now, from (2.9) we have a simple relation
$${d\,{\cal H}_{(B,\Pi,n,f)}\left(R\right)\over dR} ={n^2 +\Pi^2- B^2f^2n^2\over B\,\sqrt{n^2 + \Pi^2}}\, R + O(R^3),\eqno{(2.10)}$$
which tells us that if 
$$\left({\Pi \over n}\right)^2 < (fB)^2 - 1,\eqno{(2.11)}$$
then the energy is a decreasing function near $R =0$ and we have a local minimum at finite radius $R$.   In conclusion, the equation (2.11) is the necessary condition for the  the circular fundamental strings, D2-branes and  D0-branes to form  a stable tubular bound state.   In fact, form the condition $H'(R_*) = 0$ we can find the exact solution of  $R_*$
$$R_* =  {1\over f ~}\, \sqrt {\, {n\over \Pi}\, \,\sqrt {f^2 B^2 - 1} \,  - 1},\eqno{(2.12)}$$
This equation also shows that if the condition (2.11) is satisfied then the tube has a finite value of radius $R_*$.  

  For the fixed values of $f$ and $B$ equation (2.11) tells us that the maximum value of  $\left({\Pi/ n}\right)$ is  $\sqrt {f^2 B^2 - 1}$.  The existence of  the critical value means that there is a maximum number of F-strings which could be bound with the D0 and D2 branes to form a stable tube.   Physically, once this ratio was larger then the critical value the magnetic force from the Melvin background could not support the binding force of circular fundamental strings with D0 and D2 branes.  Thus the tube will collapse to zero radius.  This is the main result of this paper.  
  
To conclude this section we will show that the uniform magnetic field could not give a magnetic force to support the tube.   The line element  of ten-dimensional spacetimes with uniform magnetic field  is described by [7]
$$ ds^2  =  -\left (dt + \sum_{i,j=1}^{2} \epsilon_{ij}  \frac{f }{2} x^j dx^i\right)^2 + \sum_{i=1}^{2} dx^idx_i + \sum_{m=3}^{9}dx^m dx_m,  \eqno{(2.13)}$$
in which $f$ is the strength of the magnetic field and the dilaton is constant.  The BI Lagrangian (2.2) becomes
$$ {\cal L} = -\,n\,\sqrt {R^2 - E^2 + B^2}~,  \eqno{(2.14)}$$
which does not depend on the  strength of the magnetic field $f$ and thus the tube solution, if it exists, will behave as that in the flat space.  It is easy to see that the momentum conjugate to $E$ is
$$\Pi =  \,{n\,E \over \sqrt {R^2 - E^2 + B^2}~},   \eqno{(2.15)}$$ 
and the Hamiltonian density becomes
$$ {\cal H}\left(R\right)  = \sqrt {R^2  + B^2}\,\, \sqrt{n^2 + \Pi^2 },   \eqno{(2.16)}$$
which is an increasing function of $R$ and thus the tube will collapse to zero radius.  This means that the uniform magnetic field could not give a magnetic force to support the tube.

\section{Discussions}

In this paper we have used the Born-Infeld action to find the tube-like configurations in the Kaluza-Klein Melvin magnetic tube spacetime.    We have seen that the magnetic force from the Melvin background [8] could prevent 
the tubular bound configurations of  $m$ circular fundamental strings, $n$ D2-branes and $k$ D0-branes from collapsing.   However, if the ratio $m/n$ is larger then a critical value then the tube will become unstable and collapse to zero radius.  We have presented a simple analysis to find the critical value and tube radius.  We have also shown that tubular bound state of circular fundamental strings, D0 and D2 branes do not exist in the homogeneous magnetic background. 

Finally, we make following comments to conclude this paper.

1. It is known that the supersymmetry will be broken by the presence of the magnetic field as the boson and fermion will feel different force therein.  Thus the tube solutions we found in this paper are not the BPS configurations while       the tubes found in [1] had been proved to be a supersymmetric object.

2.  The results of [1-3] show that the energy of a supertube formed from  $N \times m$ straight fundamental strings and $N \times k$ D0-branes has the energy 
$${\cal H} = N|B| + N|\Pi|,  \eqno{(3.1)}$$
which is the same as the energy of $N$ supertubes in which each one is formed from $m$ straight fundamental strings and $k$ D0-branes.  Thus these configurations could be transfered to each other without costing any energy.   However, substituting the tube radius $R_*$ of (2.12) into (2.9) we see that the new tube have the energy

$${\cal H}_{(B,\Pi,n,f)} = \sqrt{B^2 + {\sqrt{f^2 B^2 -1~}\,\, n - \Pi\over f^2 \Pi}}\,\sqrt{{n\,\Pi\over \sqrt{f^2 B^2 -1~}} + \Pi ^2}.  \eqno{(3.2)}$$
\\
This relation is different from (3.1) and therefore depending on the magnitude of magnetic field $f$ (and, may be, the values of $B, \Pi , n$) a tube may split into multiple tubes (or the multiple tubes shall join into a single tube) to stabilize the system [12].   The details remain to be analyzed.

3.  As the supertube of [1-3] is supported by the angular momentum and angular momentum therein is proportional to the number of  F-string the tube radius will therefore be proportional to the number of F-string.   This property is totally different from our tube solutions.   As our tube solutions are supported by the magnetic force in the Melvin background, while {\it increasing} the number of F-string (i.e. $\Pi$) will also increase the binding force of F-strings with D0, D2 branes and thus the radius of the tube solution will {\it decrease}, as was explicitly shown in (2.12).   

4.   It is known that the tube solution in a flat space found in [1-3] is prevented from collapsing by the angular momentum, while the tube solution in a Melvin space found in this paper is prevented from collapsing by the magnetic force.     The literatures in [13,14] had shown that a circular string in a flat space could be prevented from collapsing by the angular momentum.  Therefore, it is natural to ask the question of whether the magnetic force  (and what form it shall be) could  prevent a ring-like configuration (which may be a bound state of the circular fundamental string and D-string) from collapsing ?  The interesting problem remains to be studied.

\newpage
\begin{center} {\large \bf  References} \end{center}
\begin{enumerate}
\item D. Mateos and P. K. Townsend, ``Supertubes'', Phys. Rev. Lett. 87 (2001) 011602 [hep-th/0103030]; R. Emparan, D. Mateos and P. K. Townsend, ``Supergravity Supertubes'', JHEP 0107 (2001) 011 [hep-th/0106012];; D.~Mateos, S.~Ng and P.~K.~Townsend, ``Tachyons, supertubes and brane/anti-brane systems'', JHEP  0203 (2002) 016 [hep-th/0112054].
\item D. Bak, K. M. Lee, ``Noncommutative Supersymmetric Tubes'',  Phys. Lett. B509 (2001) 168 [hep-th/0103148]; D. Bak and A. Karch, ``Supersymmetric Brane-Antibrane Configurations,'' Nucl. Phys.  B626 (2002) 165 [hep-th/011039]; D. Bak and N. Ohta, ``Supersymmetric D2-anti-D2 String,'' Phys. Lett.  B527 (2002) 131 [hep-th/0112034].
\item C. Kim, Y. Kim, O-K. Kwon, and P. Yi, ``Tachyon Tube and Supertube,''  JHEP 0309 (2003) 042 [hep-th/0307184]; Wung-Hong Huang, ``Tachyon Tube on non BPS D-branes,''  [hep-th/0407081].
\item R.C. Myers , ``Dielectric-Branes,''  JHEP 9912 (1999) 022  [hep-th/9910053].
\item M.A. Melvin, ``Pure magnetic and electric geons,'' Phys. Lett. 8 (1964) 65; G.~W.~Gibbons and D.~L.~Wiltshire, ``Space-time as a membrane in higher dimensions,'' Nucl.\ Phys.\ B287 (1987) 717 [hep-th/0109093]; G.~W.~Gibbons and K.~Maeda, ``Black holes and membranes in higher dimensional theories with dilaton fields,'' Nucl.\ Phys.\ B298 (1988) 741.
\item J.~G.~Russo and A.~A.~Tseytlin, ``Exactly solvable string models of curved space-time backgrounds,'' Nucl.\ Phys.\ B449 (1995) 91 [hep-th/9502038];``Magnetic flux tube models in superstring theory,'' Nucl.\ Phys.\ B461 (1996) 131 [hep-th/9508068].
\item  J.~G. Russo and A.~A. Tseytlin, ``Constant magnetic field in closed string
theory: An exactly solvable model,'' Nucl. Phys.  B448 (1995)  293 [hep-th/9411099]; J.~G. Russo and A.~A. Tseytlin, ``Heterotic strings in uniform magnetic field,'' Nucl. Phys. B454 (1995) 164 [hep-th/9506071]; J. David, ``Unstable magnetic fluxes in heterotic string theory,''  JHEP 0209 (2002) 006 [hep-th/0208011].
\item J.~G. Russo and A.~A. Tseytlin, ``Green-Schwarz superstring action in a curved magnetic Ramond-Ramond background,''  JHEP 9804 (1998) 014 [hep-th/9804076]; F.~Dowker, J.~P.~Gauntlett, D.~A.~Kastor and J.~Traschen, ``The decay of magnetic fields in Kaluza-Klein theory,'' Phys.\ Rev.\ D52 (1995) 6929 [hep-th/9507143]; M.~S.~Costa and M.~Gutperle, ``The Kaluza-Klein Melvin solution in M-theory,'' JHEP 0103 (2001) 027 [hep-th/0012072].
\item E. Witten, ``Bound States of Strings and p-Branes,'' Nucl.\ Phys.\ B460 (1996) 335 [hep-th/9510135].
\item  C.G.~Callan, I.R.~Klebanov,  ``D-brane Boundary State Dynamics,'' Nucl. Phys.  B465 (1996) 473-486  [hep-th/9511173].
\item  N.~Seiberg, L.~Susskind and N.~Toumbas, ``Strings in background electric field, space/time non-commutativity and a new noncritical string theory,''
JHEP 0006 (2000) 021 (2000) [hep-th/0005040]; I.R.~Klebanov and  J.~Maldacena, ``1+1 Dimensional NCOS and its U(N) Gauge Theory Dual'', I. J. Mod. Phys. A16 (2001) 922  [hep-th/0006085]; U. H. Danielsson, A. Guijosa, ans M. Kruczenski, ``IIA/B, Wound and Wrapped,'' JHEP 0010 (2000) 020 [hep-th/0009182].
\item Wung-Hong Huang, ``Condensation of Tubular D2-branes in Magnetic Field Background,'' [hep-th/0405192]
\item  J. Hoppe and H. Nicolai, ``Relativistic Minimal Surface'', Phys.Lett. B196 (1987)451.
\item  S. Frolov, A.A. Tseytlin, ``Rotating string solutions: AdS/CFT duality in non-supersymmetric sectors'', Phys.Lett. B570 (2003) 96-104 [hep-th/0306143]; D. Mateos, T. Mateos, and P. K. Townsend, ``Supersymmetry of Tensionless Rotating Strings in $AdS_5 x S^5$, and Nearly-BPS Operators'', JHEP 0312 (2003) 017 [hep-th/0309114]; R. Emparan, ``Rotating Circular Strings, and Infinite Non-Uniqueness of Black Rings'',  JHEP 0403 (2004) 064 [hep-th/0402149]

\end{enumerate}
\end{document}